\documentclass[11pt]{article}
\usepackage[dvips]{graphicx}
\usepackage{color}

\def\C{{\bf C}}

\begin{document}
\begin{center}
\bf{
NON HERMITIAN OPERATORS  WITH REAL SPECTRUM 
IN QUANTUM MECHANICS}
 \\
\vspace{0.5cm} J. da Provid\^encia
\\
{\sl \small Departamento de F\'{\i}sica,  Universidade de Coimbra, P-3004-516 Coimbra, Portugal} \\
\vspace{0.5cm}N. Bebiano\\{\sl\small Departamento de Matem\'atica,
Universidade de Coimbra, P-3004-516 Coimbra,
Portugal}\\\vspace{0.5cm}J.P. da Provid\^encia\\{\sl\small
Depatamento de F\'\i sica, Universidade da Beira Interior,
P-6201-001 Covilh\~a, Portugal}
\\
\vspace{0.5cm}\rm \today
\end{center}

\begin{abstract} Examples  are given of non-Hermitian  Hamiltonian operators
which have a real spectrum. Some of the investigated operators are
expressed in terms of the generators of the Weil-Heisenberg algebra.
{It is argued that the existence of an involutive operator $\hat J$
which renders the Hamiltonian $\hat J$-Hermitian leads to the
unambiguous definition of an} associated positive definite norm
allowing for the standard probabilistic interpretation of quantum
mechanics. Non-Hermitian extensions of the Poeschl-Teller
Hamiltonian are also considered.  Hermitian counterparts obtained by
similarity transformations are constructed.
\end{abstract}

\bigskip
\def\d{{\rm d}}
\def\e{{\rm e}}
\def\<{\langle}
\def\>{\rangle}
\section{Introduction}
In non-relativistic quantum mechanics, the Hamiltonian operator is
assumed to be Hermitian. It is well-known, however, that some
relativistic extensions, such as the Klein-Gordon theory, lead to
Hamiltonian operators $H$ which are non-Hermitian, $H\neq H^\dag$.
In such theories an indefinite norm operator $P$ occurs which
renders the Hamiltonian $P$-Hermitian, that is, $PH=H^\dag P.$
Particularly, $P$ may be the parity operator  $\cal P$ that performs
spatial reflection and has the effect $p\rightarrow -p$ and
$x\rightarrow -x.$  The indefinite norm operator $P$ does not allow
for the usual probabilistic interpretation of quantum mechanics,
because it is not positive definite. A positive definite operator
$Q$ which can play the role of $P$ in the sense that $QH=H^\dag Q$,
is still needed for the conventional interpretation of quantum
mechanics.

Non-Hermitian Hamiltonian operators  with a real spectrum have been
the object of intense research activity \cite{scholtz,[1],[2],[3]}.
For instance, the Hamiltonian $H=p^2+x^2+ix^3$, has been studied by
Bender and others \cite{[1]}, who observed that its spectrum is real
due to the $\cal P\cal T$-symmetry being unbroken. That operator is
not symmetric under $\cal P$ or $\cal T$ separately, but is
invariant under their combined operation and so it is said to
possess space-time symmetry. Here, $\cal T$ denotes the anti-linear
time-reversal operator, which has the effect $p\rightarrow -p$,
$x\rightarrow x$ and $i\rightarrow -i$. Following Bender's work,
many researchers, as, for instance, Gonz\'alez Lop\'ez, \cite{[2]}
investigated such non-Hermitian Hamiltonians with real spectra.

The construction of positive norm operators required by the quantum
mechanical probabilistic interpretation of non-Hermitian
Hamiltonians of this type is a topic of current interest. The
problem of non-uniqueness of the metric was addressed by Scholtz
{\it et al} \cite{scholtz} who resolved it by considering an
irreducible set of observables. Bender {\it at al} \cite{[1]}, in
the context of $\cal P\cal T$-invariant theories, defined a new
operator $\cal\,C$ and the $\cal C\cal P\cal T$ scalar product,
which is positive definite. However, to construct $\cal C$ the
eigenvalues and eigenvectors of the Hamiltonian have to be
determined, which can be done explicitly for soluble models but in
general situations only a perturbative expansion of $\cal C$ is
available. (For extensive bibliography, see the references cited in
\cite{[3]}.)

Let the non-Hermitian Hamiltonian $H$ acting on the Hilbert space
$\cal\,H$ have real eigenvalues $\lambda_j$ and let the
corresponding right and left eigenvectors, denoted by $|\phi_j\>$
and $\<\psi_j|$, respectively, form two  complete systems. We have
$$H|\phi_j\>=\lambda_j|\phi_j\>,\quad \<\psi_j| H=\lambda_j\<\psi|.
$$ The completeness of the eigenvectors $|\phi_i\>$ means that any
$|\xi\>\in \cal\,H$ may be expanded as $|\xi\>=\sum_{i=1}^n
|\phi_i\> c_i,$ for certain  $c_i\in {\bf C }$. Mostafazadeh
\cite{mostafa} has shown that there exists  positive definite
Hermitian operators $Q$ such that $QH$ is Hermitian, $QH=H^\dag Q$,
and under the similarity transformation
$$\tilde H=Q^{1\over2}HQ^{-{1\over2}},
$$
a Hermitian operator $\tilde H$ is obtained.

Mostafazadeh's result is easily understood in the finite dimensional
case. Let ${\cal H}=\C^n$ and denote by $\langle\psi|\phi\rangle $
the inner product of the vectors $|\phi\>,|\psi\> \in {\C}^n$.
Dirac's {\it bra-ket} notation is used throughout. Let $A$ be an
$n\times n$ non-Hermitian matrix, that is $A\neq A^\dag$, where
$A^\dag$ denotes the adjoint matrix. Let us also assume that the
eigenvalues $\lambda_j$ of $A$ are real and that the right and left
eigenvectors, denoted, respectively, by $|\phi_j\rangle$ and
$\langle\psi_j|$, form two complete systems,
$$A|\phi_j\rangle=\lambda_j|\phi_j\rangle,\quad \langle\psi_j|A=\lambda_j\langle\psi_j|.
$$
Assuming for simplicity that the eigenvalues are distinct, it
follows that $\langle\psi_j|\phi_k\rangle=0$ if $j\neq k.$ If the
eigenvalues are conveniently normalized, we have moreover
$$\langle\psi_j|\phi_k\rangle=\delta_{jk}.
$$
Let $Q$ be the operator such that $|\psi_j\rangle=Q|\phi_j\rangle$.
This operator is Hermitian and positive definite (notice that
$\langle\psi_j|\phi_k\rangle=\langle\phi_j|\psi_k\rangle$) and has
the property that $QA$ is Hermitian, $QA=A^\dag Q$. These assertions
may be easily verified. Thus, under the similarity transformation
$$\tilde A=Q^{1\over2}AQ^{-{1\over2}},$$
a Hermitian operator is obtained, in agreement with Mostafazadeh's
result \cite{mostafa}. The operator $Q$ induces the inner product
$$\langle\psi| Q|\phi\rangle,$$ satisfying $ \langle\psi| Q|\psi\rangle>0,$ for $
|\psi\rangle\neq0 $. Given an arbitrary vector $|\xi\rangle\in{\bf
C}^n$ such that $\langle\xi|Q|\xi\rangle=1$, according to the rules
of quantum mechanics, the component
$c_j=\langle\phi_j|Q|\xi\rangle=\langle\psi_j|\xi\rangle$ of the
vector $|\xi\rangle=\sum c_j|\phi_j\rangle$ may be understood as the
amplitude (square root) of the probability for the result of a
measurement of the observable $A$ to be $\lambda_j$. However, $Q$
must be fixed by some supplementary requirement. Physical theories
involving non-Hermitian Hamiltonians with real eigenvalues also
prescribes certain Hermitian involutive operators $J,\;J^2=I$, such
that $JH$ is Hermitian, $JH=H^\dag J$, so that the eigenvectors
$|\phi_j\>$ of $H$ may be normalized according to
$\langle\phi_i|J|\phi_j\rangle=\eta_i\delta_{ij},\;\eta_i=\pm1$.
Since $\langle\phi_j|J$ is a left eigenvector, i.e.,
$\langle\phi_j|JH=\lambda_j\langle\phi_j|J$, it is natural to chose
$\langle\psi_j|=\eta_j\langle\phi_j|J$ as the desired left
eigenvector satisfying $\langle\psi_j|\phi_j\rangle=1$ and to define
$Q$ by the relation
$|\psi_j\rangle=\eta_jJ|\phi_j\rangle=Q|\phi_j\rangle$. By this
prescription, $Q$ is fixed unambiguously  and the relation
$JQJ=Q^{-1}$ is satisfied, as the following argument shows. Indeed,
let us consider the left eigenvectors of $H$,
$\langle\psi_i|=\langle\phi_i|Q=\eta_i\langle\phi_i|J.$ Since
$$|\psi_i\rangle=\eta_iJ|\phi_i\rangle=\eta_iJQ^{-1}|\psi_i\rangle
=Q|\phi_i\rangle=Q(\eta_iJ|\psi_i\rangle)=\eta_iQJ|\psi_i\rangle,$$
we have $QJ|\psi_i\rangle=JQ^{-1}|\psi_i\rangle$. Since, moreover,
the $|\psi_i\rangle$ constitute a complete system of vectors, the
result follows. This conclusion is in agreement with the criterium
proposed by Bender and collaborators \cite{bender1} for the complete
specification of a suitable norm operator $Q$, namely, $J\log Q
J=-\log Q$.

We  investigate simple examples of non-Hermitian  Hamiltonians which
have real spectra and may be diagonalized with the help of algebraic
methods. Moreover, these operators have complete systems of right
and left eigenvectors.

\section{The harmonic oscillator}
For the sake of completeness, let us consider the harmonic
oscillator Hamiltonian \begin{eqnarray}H={1\over2}(p^2+x^2),
\end{eqnarray} which acts on the space $L^2$ of square integrable
differentiable functions of the real variable $x$ endowed with
 the usual inner product
$$\langle\phi|\psi\rangle=\int\d x\,\phi(x)^*\psi(x).
$$
As it is well-known,   $p:L^2\rightarrow L^2$ is the differential
operator $f(x) \rightarrow-i(\d f/\d x)$ and $x:L^2\rightarrow L^2$
is the multiplicative operator $f(x)\rightarrow xf(x)$. These
operators satisfy the {\it quantum condition $[p,x]=-i$} and the
harmonic oscillator Hamiltonian  is Hermitian. Following the well
known Dirac's approach, its spectrum is determined with the help of
the Weil-Heisenberg algebra generated by the {\it creation} and {\it
annihilation operators} $a^\dag$ and $a$, respectively, defined by
the linear combinations
\begin{equation}a={1\over\sqrt{2}}(x+ip),\quad
a^{\dag}={1\over\sqrt{2}}(x-ip).\label{ho}
\end{equation}
As the notation conveys, $ a^{\dag}$ is the adjoint of $a$ and these
operators satisfy the commutation relation $[a,a^{\dag}]=1$.  One
easily finds
$$H=a^{\dag}a+{1\over2}.
$$
Moreover, for the vector $|\phi_0\rangle$ such that
$a|\phi_0\rangle=0$ and $|\phi_n\rangle=a^{{\dag}n}|\phi_0\rangle,$ we have
$$H|\phi_n\rangle=\left(n+{1\over2}\right)|\phi_n\rangle,\quad
n=0,1,2,\cdots.
$$
The vector $|\phi_0\rangle$ stands, in Dirac's notation, for nothing
else than a solution $\phi_0(x)$  of the differential equation
$$\left(x+{\d\over \d x}\right)\phi_0(x)=0,\quad \phi_0(x)=K_0\e^{-{x^2\over2}}.$$ On the other hand
$|\phi_n\rangle$ is identified with the function
\begin{eqnarray}\phi_n(x)=K_n\left(x-{\d\over \d
x}\right)^n\e^{-{x^2\over2}},\label{howf}\end{eqnarray} where $K_n$
is a normalization constant.

\section{Non-Hermitian operators with a real spectrum}
We give simple examples of non-Hermitian operators with a real
spectrum and a bi-orthogonal set of eigenvectors. {In each case, we
show that there exists an involutive operator $\cal J$ which renders
the Hamiltonian $\cal J$-Hermitian and allows for the unambiguous
definition of a positive definite norm operator suitable for the
conventional quantum mechanical interpretation.}
\bigskip
\subsection{The extended harmonic oscillator}
\bigskip
We begin by considering the operator
\begin{eqnarray}H_\beta={\beta\over2}(p^2+x^2)+i{\sqrt{2}}\,p=\beta
a^{\dag}a +(a-a^\dag)+{\beta\over2},\quad \beta>0. \end{eqnarray}
Although $H_\beta$ is non-Hermitian, and it is even not ${\cal
P}\cal T$-symmetric, it is nevertheless ${\cal P}$-Hermitian, i.e.,
${\cal P}H_\beta=H_\beta^\dag{\cal P}$. In order to determine the
spectrum of $H_\beta$,
we write
$$H_\beta=\beta\left(a^\dag+{1\over\beta}\right)\left(a-{1\over\beta}\right)+{1\over\beta}+{\beta\over2}.
$$
The spectrum and the eigenvectors of $H_\beta$ are easily determined
by the usual technique due to Dirac. Although the operators
$(a^\dag+{1/\beta})$ and $(a-{1/\beta})$ are not the adjoint of each
other, they generate a Weil-Heisenberg algebra, so we still have
$$\left[\left(a^\dag+{1\over\beta}\right)\left(a-{1\over\beta}\right),
\left(a^\dag+{1\over\beta}\right)^n\right]=n\left(a^\dag+{1\over\beta}\right)^n,
$$
$$\left[\left(a^\dag+{1\over\beta}\right)\left(a-{1\over\beta}\right),
\left(a-{1\over\beta}\right)^n\right]=-n\left(a-{1\over\beta}\right)^n.
$$
It follows that
$$H_\beta|R_n\rangle=\left({1\over\beta}+\beta\left(n+{1\over2}\right)\right)|R_n\rangle,
$$
where
$$\left(a-{1\over\beta}\right)|R_0\rangle=0,\quad
|R_n\rangle=\left(a^*+{1\over\beta}\right)^n|R_o\rangle,
$$
which fixes the  right eigenvectors. The  left eigenvectors are
given by
$$\langle L_n|H_\beta=\left({1\over\beta}+\beta\left(n+{1\over2}\right)\right)\langle L_n|,
$$
where
$$\langle L_n|\left(a^\dag+{1\over\beta}\right)=0,\quad
\langle L_n|=\langle L_0|\left(a-{1\over\beta}\right)^n.
$$
The spectrum of $H_\beta$ is given by
$$\sigma(H_{\beta})={1\over\beta}+
\beta\left\{{1\over2},{3\over2},{5\over2},\cdots\right\}.
$$
We have  assumed $\beta\neq0,$ so the spectrum is obviously real.
The right eigenvectors and the left eigenvectors are represented,
respectively, by the functions
$$R_n(x)=K_n\left(x+{\sqrt{2}\over\beta}-{\d\over\d
x}\right)^n\exp\left(-{1\over2}\left(x-{\sqrt{2}\over\beta}\right)^2\right)
$$
and
$$L_n(x)=K_n\left(x-{\sqrt{2}\over\beta}-{\d\over\d
x}\right)^n\exp\left(-{1\over2}\left(x+{\sqrt{2}\over\beta}\right)^2\right),
$$
where $K_n$ denotes a normalization constant.  Obviously,
$L_n(x)=(-1)^nR_n(-x),$ {that is, the left eigenvectors are
essentially the involution of the right eigenvectors generated by
$\cal P$.} For
$$K_n^2={\sqrt{\pi}\over2^n}\exp(-\beta^{-2})
$$
the eigenfunctions are orthonormal, $\langle
L_n|R_m\rangle=\delta_{nm}.$

It is clear that the operators $(a^{\dag}+1)/\beta,(a-1)/\beta$ are
related to the operators $a^{\dag},a$ by a similarity
transformation,
$$a^{\dag}+{1\over\beta}=\e^{(a^{{\dag}}+a)/\beta}\,a^\dag\,\e^{-(a^{{\dag}}+a)/\beta},\quad
a-{1\over\beta}=\e^{(a^{{\dag}}+a)/\beta}\,a\,\e^{-(a^{{\dag}}+a)/\beta}.
$$
Thus, the operator
$$\widetilde H_\beta=\e^{-(a^{{\dag}}+a)/\beta}\,H_\beta\,\e^{(a^{{\dag}}+a)/\beta}
$$
is Hermitian. Moreover,
$|L_n\rangle=\e^{-2(a^{{\dag}}+a)/\beta}\,|R_n\rangle$ and $\langle
R_n| \e^{-2(a^{{\dag}}+a)/\beta} |R_m\rangle=\delta_{nm}$.

Other operators share with $Q$ the property of rendering $H_\beta$
Hermitian. With respect to the orthonormal basis constituted by the
eigenvectors $|\phi_n\rangle$ of the harmonic oscillator
(\ref{howf}), the operator $H_\beta$ is represented by the
tridiagonal matrix
$$M_\beta=\pmatrix{{\beta/2}&\sqrt{1}&0&0&\cdots\cr
-\sqrt{1}&{3\beta/2}&\sqrt{2}&0&\cdots\cr
0&-\sqrt{2}&{5\beta/2}&\sqrt{3}&\cdots\cr
0&0&-\sqrt{3}&7\beta/2&\cdots\cr
\vdots&\vdots&\vdots&\vdots&\ddots&}.
$$
 The matrix $M_\beta$ is $J$-Hermitian for $J={\rm
diag}(1,-1,1,-1,\cdots)$. Let $\cal J$ denote the operator which is
represented by the matrix $J$ with respect to the basis constituted
by the eigenvectors $|\phi_n\rangle$. In the present case, $\cal J$
is precisely the parity operator $\cal P$. If $|R_n\rangle$ is a
right eigenvector of $H_\beta$, then $\langle R_n|{\cal J}$ is a
left eigenvector. For a vector $|\Xi\rangle$ normalized according to
$\langle\Xi|Q|\Xi\rangle=1$, where  $Q=\e^{-2(a^{{\dag}}+a)/\beta}$,
and for the eigenvectors of $H_\beta$ normalized according to
$\langle R_i|{\cal J}|R_j\rangle=(-1)^i\delta_{ij}$ , the quantity
$|\langle R_i|{\cal J}|\Xi\rangle|^2$ has the meaning of a
probability.

We observe that ${\hat J}(\log Q){\cal J}=-\log Q$, as may be easily
checked, keeping in mind that $\log Q={-2(a^{{\dag}}+a)/\beta}.$
This is in agreement with Bender's \cite{bender1} criterium for the
complete specification of the norm operator $Q$.
\subsection{The Swanson Hamiltonian}
Next, we consider an operator of a class which has been proposed by
Swanson 
\cite{swanson,jones,musumbu}, namely the non-Hermitian Hamiltonian
with a real spectrum
\begin{eqnarray}H_{\theta}={1\over2}(p^2+x^2)
-{i\over2}\tan2\theta(p^2-x^2),\end{eqnarray} where $\theta$ is a
real parameter, $ -{\pi\over4}<\theta<{\pi\over4} $. For the
operators $a^{\dag},a$ of eq. (\ref{ho}) we also have
\begin{eqnarray}H_{\theta}=
a^{\dag}a+i{\tan2\theta\over2}\left( a^{{\dag}2}+
a^2\right)+{1\over2}\nonumber.
\end{eqnarray}
In terms of the differential operators
\begin{equation}\label{eho} c=\cos\theta\, a+i~{\sin\theta}\,a^{\dag},
\quad\nonumber c^{\ddag}=\cos\theta\, a^{\dag}+i~\sin\theta\,a
\end{equation} a diagonal, oscillator-like form, is obtained,
$$H_{\theta}=
\omega\left(c^{\ddag}c+{1\over2}\right),\quad
\omega={1\over\cos2\theta}.
$$
The operators $c,c^{\ddag}$ satisfy $[c,c^{\ddag}]=1$. Although
$c^{\dag}\neq c^{\ddag}$, these operators realize a certain
representation of the Weil-Heisenberg algebra.
It follows that the spectrum is
$$\sigma(H_{\theta})=\omega\left\{{1\over2},{3\over2},{5\over2},\cdots\right\}.
$$
Keeping in mind the explicit expressions of the differential
operators (\ref{eho})
\begin{eqnarray}&&c={
\e^{i\theta}\over\sqrt{2}}\,
x+\,{
\e^{-i\theta}\over\sqrt{2}}\,{\d\over\d x},\quad 
c^{\ddag}={
\e^{i\theta}\over\sqrt{2}}\, x
\,-{
\e^{-i\theta}\over\sqrt{2}}\,{\d\over\d x},\nonumber
\end{eqnarray}
the left and right eigenvectors are easily obtained. The right
eigenvectors  satisfy
$$H_{\theta}|R_n\rangle=\left({1\over2}+n\right)\omega|R_n\rangle
$$
and are given by
$$c|R_0\rangle=0,\quad |R_n\rangle=K_n(c^{\ddag})^n|R_0\rangle,
$$
where $K_n$ is a real normalization constant. The left eigenvectors
satisfy
$$\langle L_n|H_{\theta}=\left({1\over2}+n\right)\omega\langle L_n|
$$
and are given by
$$\langle L_0|c^{\ddag}=0,\quad \langle L_n|=K_n\langle L_0|c^n.
$$
The last equation is equivalent to
$$(c^{\ddag})^\dag|L_0\rangle=0,\quad |L_n\rangle=K_n(c^{\dag})^n|L_0\rangle.
$$
As an example, we present the explicit expressions of the lowest
right and left eigenvectors, $$R_0(x)=K_0\exp \left(-{x^2\over2}~
\e^{2i\theta}\right),\quad L_0(x)=K_0\exp \left(-{x^2\over2}~
\e^{-2i\theta}\right).
$$
For an appropriate value of the normalization constant $K_n$, the
eigenfunctions are orthonormal, $\langle
L_n|R_m\rangle=\delta_{nm}.$

It is clear that the operators $c^{\ddag},c$ are related to the
operators $a^{\dag},a$ by a similarity transformation,
$$c^{\ddag}=\e^{i{\theta\over2}(a^{2}-a^{\dag 2})}\,a^\dag\,
\e^{-i{\theta\over2}(a^{2}-a^{\dag 2})},\quad
c=\e^{i{\theta\over2}(a^{2}-a^{\dag2})}\,a\,
\e^{-i{\theta\over2}(a^{2}-a^{\dag2})},
$$
Thus, the Hermitian operator
$$\widetilde H_{\theta}=
\e^{-i{\theta\over2}(a^{2}-a^{\dag2})}\,H_\theta\,
\e^{i{\theta\over2}(a^{2}-a^{\dag2})}=\omega\left({1\over2}+a^{\dag}a\right)
$$
is the harmonic oscillator operator with frequency $\omega$, as
observed by Jones \cite{jones}. Moreover,
$$|L_n\rangle=\e^{-i{\theta}(a^{2}-a^{\dag2})}\,|R_n\rangle$$ and
$$\langle R_n|\e^{-i{\theta}(a^{2}-a^{\dag2})} |R_m\rangle=\delta_{nm}.$$
For a vector $|\Xi\rangle$ normalized according to
$\langle\Xi|Q|\Xi\rangle=1$, where
$$Q=\e^{-i{\theta}(a^{2}-a^{\dag2})},$$ the quantity $|\langle
R_i|Q|\Xi\rangle|^2=|\langle L_i|\Xi\rangle|^2$ has the meaning of a
probability. We observe that the knowledge of the relative
probability does not require the determination of the full set of
left- and right-eigenfunctions, but only of the eigenfunctions
related to the relevant transitions. Our conclusions are similar to
those of Ref. \cite{jones}. However, our approach is different,
since it is based on the explicit determination of the left- and
right-eigenfunctions.

Next we will show that there exits a Hermitian involutive operator
$\cal J$ (such that ${\cal J}^2=Id$) which also renders $H_{\theta}$
Hermitian, that is, ${\cal J}H_{\theta}=H_{\theta}^\dag{\cal J}$.
With respect to the orthonormal basis constituted by the
eigenvectors $|\phi_n\rangle$ of the harmonic oscillator
(\ref{howf}), $H_\theta$ is represented by the matrix
$$M_\theta=\pmatrix{{1/2}&0&i\beta\sqrt{1.2}&0&0&\cdots\cr
0&{3/2}&0&i\beta\sqrt{2.3}&0&\cdots\cr
i\beta\sqrt{1.2}&0&{5/2}&0&i\beta\sqrt{3.4}&\cdots\cr
0&i\beta\sqrt{2.3}&0&7/2&0&\cdots\cr
0&0&i\beta\sqrt{3.4}&0&9/2&\cdots\cr
\vdots&\vdots&\vdots&\vdots&\vdots&\ddots&},\quad
\beta={\tan2\theta\over2}.
$$
It is clear that ${J}M_\theta=M_\theta^\dag{J}$ for
$J=I_2\oplus-I_2\oplus I_2\oplus-I_2\oplus\cdots$. {Let $\cal J$
denote the operator which is represented by the matrix $J$ with
respect to the system of the eigenvectors $|\phi_n\rangle$
(\ref{howf}). It follows that ${\cal J}H_\theta=H_\theta^\dag{\cal
J}$.

If $|R_n\rangle$ is a right eigenvector of $H_{\theta}$, then
$\langle R_n|{\cal J}$ is a left eigenvector, so that $\cal J$ is
the required involutive operator. For a vector $|\Xi\rangle$
normalized according to $\langle\Xi|Q|\Xi\rangle=1$, and for
eigenvectors of $H_{\theta}$ normalized according to $\langle
R_i|{\cal J}|R_j\rangle=\eta_i\delta_{ij}$, where $\eta_i=1$ for
$i=4n,~(4n+1),\,n=0,1,2,\cdots$ and $\eta_i=-1$ for
$i=(4n-2),~(4n-1),\,n=1,2,\cdots,$ the quantity $|\langle R_i|{\cal
J}|\Xi\rangle|^2$ has the meaning of a probability.} We observe that
${\cal J}(\log Q){\cal J}=-\log Q$, for $\log
Q={-i\theta(a^{2}-a^{\dag2})}$, as may be easily checked. This is in
agreement with Bender's \cite{bender1} criterium for the complete
specification of the norm operator $Q$.

\subsection{Non-Hermitian extensions of the Poeschl-Teller Hamiltonian} Let us
now consider the Poeschl-Teller Hamiltonian \cite{fluegge}
\begin{eqnarray}\label{PT}
H=p^2-{\gamma(\gamma-1)\over\cosh^2x},\quad\gamma>1.
\end{eqnarray}
In terms of the operators $$A^\dag=-ip+(\gamma-1)\tanh~x,\quad
A=ip+(\gamma-1)\tanh~x, $$ we obtain $$H=A^\dag A-(\gamma-1)^2.
$$
For a detailed algebraic treatment of the Poeschl-Teller
Hamiltonian, see \cite{plyushchay}. The discrete spectrum of this
Hamiltonian is easily determined. It is the the set of eigenvalues:
$$E_n=-(\gamma-1-n)^2, ~~~ n=0,1,\cdots\leq\gamma-1.$$

Next we investigate non-Hermitian extensions of the Poeschl-Teller
Hamiltonian \cite{ahmed}. To start with, we consider the $\cal
PT$-symmetric extension, obtained from (\ref{PT}) through the
replacement $x\rightarrow x-i\alpha$,
\begin{eqnarray}
&H_\alpha&=p^2-{\gamma(\gamma-1)\over\cosh^2(x-i\alpha)},\nonumber\\
&&=p^2-{2\gamma(\gamma-1)(1+\cosh
2x\cos2\alpha+i\sinh2x\sin2\alpha)\over(\cosh 2x-\cos2\alpha)^2},
\quad\gamma>1,\label{PT-PT}
\end{eqnarray} where $\alpha$ denotes a real parameter. The
Hamiltonian (\ref{PT-PT}) is $\cal P$-Hermitian: ${\cal
P}H_\alpha=H_\alpha^\dag{\cal P}.$ In terms of the operators
$$A_\alpha^\ddag=-ip+(\gamma-1)\tanh(x-i\alpha),\quad A_\alpha=ip+(\gamma-1)\tanh(x-i\alpha), $$ we
obtain $$H_\alpha=A_\alpha^\ddag A_\alpha-(\gamma-1)^2.
$$ The spectrum of this Hamiltonian is real, and coincides
with the spectrum of the Hamiltonian (\ref{PT}), as may be easily
seen following the technique described in \cite{plyushchay}. This is
not surprising, because $H_\alpha$ is similar to $H$,
$$H_\alpha=\e^{-\alpha p}H\e^{\alpha p}.
$$
Moreover, $H_\alpha$ is $Q$-Hermitian, for $Q=\e^{2\alpha p}$, i.e.,
$QH_\alpha=H_\alpha^\dag Q$. The norm defined with the help of the
positive definite operator $Q$ is appropriate for the usual
statistical interpretation of quantum-mechanics.

Finally we consider the non-$\cal PT$-symmetric extension of the
Poeschl-Teller Hamiltonian, obtained from (\ref{PT}) through the
replacement $p\rightarrow p~\e^{i\theta},~x\rightarrow
x~\e^{-i\theta}$.
\begin{eqnarray}
&H_\theta&=(\eta+i\zeta)^2~p^2-{\gamma(\gamma-1)\over\cosh^2
((\eta-i\zeta) x)},\nonumber\\
&&=(\eta^2-\zeta^2 +2i\eta\zeta)~p^2-{2\gamma(\gamma-1)(1+\cosh
(2\eta x)~\cos(2\zeta x)+i\sinh(2\eta x)~\sin(2\zeta x))\over(\cosh
(2\eta x)-\cos(2\zeta x))^2 }~,\nonumber\\&&
\gamma>1,~\eta=\cos\theta,~\zeta=\sin\theta,
~-{\pi\over4}<\theta<{\pi\over4}\label{PT-n-PT}.
\end{eqnarray}
In terms of the operators
$$A_\theta^\ddag=-ip\e^{i\theta}+(\gamma-1)\tanh(x\e^{-i\theta}),
\quad A_\theta=ip\e^{i\theta}+(\gamma-1)\tanh(x\e^{-i\theta}), $$ we
obtain $$H_\theta=A_\theta^\ddag A_\theta-(\gamma-1)^2.
$$
The spectrum of the Hamiltonian (\ref{PT-n-PT}) is real, and
coincides with the spectrum of the Poeschl-Teller Hamiltonian, as
may be easily seen following the technique described in
\cite{plyushchay}. This is not surprising, because $H_\theta$ is
similar to $H$,
$$H_\theta=\e^{-{\theta\over2}(px+xp)}H\e^{{\theta\over2} (px+xp)}.
$$
Moreover, $H_\theta$ is $Q$-Hermitian, for $Q=\e^{\theta( px+xp)}$,
i.e., $QH_\theta=H_\theta^\dag Q$, so that the norm defined with the
help of the positive definite operator $Q$ is appropriate for the
conventional quantum-mechanical interpretation.
\section{Conclusions}
We have presented simple examples of non-Hermitian models in quantum
mechanics for which the Hamiltonian $H$ has a real spectrum. An
involutive ${\cal J}$ operator such that $H^\dag={\cal J}H{\cal J}$
is identified and explicitly constructed in each case. The operator
$\cal J$ allows for the definition of an indefinite norm, which,
obviously is not suitable for the standard quantum mechanical
interpretation. We discuss the important role played by this
operator in allowing for the unambiguous  definition of a positive
definite norm operator $Q$, suitable for the usual quantum
mechanical interpretation. In indefinite inner product spaces, $\cal
J$-Hermitian operators are known to have a spectrum which is
symmetric relatively to the real axis or a real spectrum.

\end{document}